\documentclass[12pt,aps,prb,preprint]{revtex4}
\begin{document}
\title{

Comment on ``On an identity for the volume integral of the square
of a vector field" }
\author{Robert L. Intemann}
\affiliation{Temple University, Department of Physics,
Philadelphia, PA 19122} \email{intemann@temple.edu}
\date{\today}

\maketitle

In a paper on the significance of the vector potential squared,
Gubarev, Stodolsky and Zakharov\cite{gubarev} quote the following
identity which holds for vector fields which vanish sufficiently
rapidly at infinity that surface terms make no contribution:
\begin{equation}
\int d^3x {\bf A(x)}^2=\int d^3x\int d^3x^{\prime}\frac{{\bf
\nabla}\cdot{\bf A(x)}{\bf\nabla^{\prime}}\cdot{\bf
A(x^{\prime})}+{\bf\nabla}\times{\bf
A(x)}\cdot{\bf\nabla^{\prime}}\times{\bf
A(x^{\prime})}}{4\pi\mid{\bf x}-{\bf x^{\prime}}\mid}. \label{one}
\end{equation}
They offer no proof of the identity other than to remark in a
footnote that the relationship can be established by transforming
to position space the simple vector identity $({\bf k}\times{\bf
a})^2=k^2{\bf a}^2-({\bf k}\cdot{\bf a})^2$ satisfied by ${\bf
a(k)}$, the momentum-space counter part (i.e., the Fourier
transform) of ${\bf A(x)}$.

Recently, Stewart\cite{stewart} has provided a position-space
derivation of Eq.\ (\ref{one}) based on a Helmholtz separation of
${\bf A(x)}$ into its longitudinal and transverse contributions.
In a comment on Stewart's paper, Durand\cite{durand} offers
another position-space derivation, one that employs only basic
operations of vector analysis and the solution to Poisson's
equation for a point charge.

In this note, we provide the derivation that appears to have been
alluded to by Gubarev, Stodolsky and Zakharov.  It makes use of
the Fourier transform along with basic techniques of vector
analysis and some elementary integration, but does not require the
solution of any differential equations. As noted by
Durand,\cite{durand} this identity can be generalized to the case
where the integrand of the volume integral involves the scalar
product of two different vector fields, both of which diminish
sufficiently rapidly at infinity.
 Since it is no more difficult to do so, we shall treat this more
 general case.\cite{generalization}

 So, let ${\bf A}({\bf x})$ and ${\bf B}({\bf x})$
 be two complex vector fields that are well behaved (i.e., that
 satisfy the Dirichlet conditions) and vanish sufficiently rapidly
 at infinity that there is no contribution from any surface terms.
  Then ${\bf A}({\bf x})$ may be represented by the Fourier
  integral,
 \begin{equation}
  {\bf A}({\bf x})=\int  d^3k\, {\bf a}({\bf k})\,
 e^{i{\bf k}\cdot{\bf x}}
 \end{equation}
  with its Fourier transform given by
\begin{equation}
{\bf a}({\bf k})=\frac{1}{(2\pi)^3}\int d^3x\,{\bf A}({\bf x})\,
e^{-i{\bf k}\cdot{\bf x}},
\end{equation}
 and with a similar pair of relations holding between ${\bf B}({\bf
 x})$ and its Fourier transform ${\bf b}({\bf k})$.
 With these representations, the integral of interest becomes,\cite{complex}
 \begin{eqnarray}
 \int d^3x\,{\bf A(x)}\cdot{\bf B^{\ast}(x)} & = & \int\,
 d^3k\int d^3k^{\prime}\,{\bf a(k)}\cdot{\bf
 b^{\ast}(k^{\prime})}\int d^3x\,e^{i({\bf k-k^{\prime}})\cdot {\bf
 x}}\label{four}\\
    & = & (2\pi)^3\int\,d^3k\,{\bf a(k)}\cdot{\bf b^{\ast}(k)},
    \label{five}
 \end{eqnarray}
 where, in obtaining the last line, we have used the standard Fourier integral representation for the
 Dirac delta function,
 \[\delta({\bf k})=\frac{1}{(2\pi)^3}\int\,d^3x\, e^{i{\bf
 k}\cdot{\bf x}},\]
 along with its basic properties.

 Employing the familiar identity
 ${\bf k}\times({\bf k}\times{\bf a})={\bf k}({\bf k}\cdot{\bf a})-({\bf k}\cdot{\bf k}){\bf
 a}$, we may separate a vector field into its transverse and
 longitudinal parts,
 \begin{equation}
 {\bf a}=\frac{{\bf k}({\bf k}\cdot{\bf a})}{k^2}-\frac{{\bf k}\times({\bf k}\times{\bf
 a})}{k^2},
 \end{equation}
 and, with the aid of this decomposition, rewrite Eq.\ (\ref{five}) as
 \begin{eqnarray}
 \int d^3x{\bf A(x)}\cdot{\bf
 B^{\ast}(x)} & = & (2\pi)^3\!\!\int d^3k\left[\frac{{\bf k}({\bf k}\cdot
 {\bf a})}{k^2}-\frac{{\bf k}\times({\bf k}\times{\bf
 a})}{k^2}\right]\!\!\cdot\!\!
 \left[\frac{{\bf k}({\bf k}\cdot
 {\bf b^{\ast}})}{k^2}-\frac{{\bf k}\times({\bf k}\times{\bf
 b^{\ast}})}{k^2}\right]\\
    & = & (2\pi)^3\!\!\int d^3k\left[\frac{({\bf k}\cdot{\bf a})({\bf k}\cdot{\bf
    b^{\ast}})}{k^2}+\frac{({\bf k}\times{\bf a})\cdot({\bf k}\times{\bf
    b^{\ast}})}{k^2}\right],
    \label{eight}
 \end{eqnarray}
 where some basic vector identities have been employed in obtaining Eq.\ (\ref{eight}).  Let us now transform back to coordinate space. Starting with the
 first term on the right hand side of Eq.\ (\ref{eight}), we have
 \begin{eqnarray}
 \int\frac{d^3k}{k^2}({\bf k}\cdot{\bf a})({\bf
 k}\cdot{\bf b^{\ast}}) & = &
 \frac{1}{(2\pi)^6}\int\frac{d^3k}{k^2}\int d^3x{\bf k}\cdot{\bf
 A(x)}e^{-i{\bf k}\cdot{\bf x}}\int d^3x^{\prime}{\bf
 k}\cdot{\bf B^{\ast}(x^{\prime}})e^{i{\bf k}\cdot{\bf
 x^{\prime}}}\\
   & = & \frac{1}{(2\pi)^6}\int\frac{d^3k}{k^2}\int d^3x{\bf
   A(x)}\cdot{\bf\nabla}e^{-i{\bf k}\cdot{\bf x}}\int
   d^3x^{\prime}{\bf B^{\ast}(x^{\prime})}\cdot{\bf\nabla^{\prime}}e^{i{\bf k}\cdot{\bf
 x^{\prime}}}\\
   & = & \frac{1}{(2\pi)^6}\int\frac{d^3k}{k^2}\int
   d^3x{\bf\nabla}\cdot{\bf A(x)}\int d^3x^{\prime}{\bf\nabla^{\prime}}\cdot{\bf
   B^{\ast}(x^{\prime})}e^{i{\bf k}\cdot({\bf x^{\prime}}-{\bf
   x})}\label{eleven}\\
     & = & \frac{1}{(2\pi)^3}\int d^3x\int d^3x^{\prime}\frac{{\bf\nabla}\cdot{\bf A(x)}
     \,{\bf\nabla^{\prime}}\cdot{\bf B^{\ast}(x^{\prime})}}{4\pi\mid{\bf x}-{\bf x^{\prime}}\mid},
   \label{twelve}
 \end{eqnarray}
 where, in obtaining Eq.\ (\ref{eleven}), we have integrated by parts, applied the divergence
   theorem, and assumed that the resulting surface term makes no
   contribution.  To obtain Eq.\ (\ref{twelve}), we have performed
   the integration over momentum space using spherical coordinates
   as follows:
   \[\int \frac{d^3k}{k^2}e^{i{\bf k}\cdot{\bf R}}=2\pi\int_{0}^{\infty}
   dk\int_{-1}^{1}d\mu e^{ikR\mu}=\frac{4\pi}{R}\int_{0}^{\infty}
   dk\frac{\sin(kR)}{k}=\frac{4\pi}{R}\int_{0}^{\infty} dx\frac{\sin x}{x}=
   \frac{(2\pi)^3}{4\pi R}.\]

 The second term on the right hand side of Eq.\ (\ref{eight}) is
 transformed to coordinate space in a similar manner.
 \begin{eqnarray}
 \int\frac{d^3k}{k^2}({\bf k}\times{\bf a})\cdot({\bf k}\times{\bf
 b^{\ast}}) & = &
 \frac{1}{(2\pi)^6}\int\frac{d^3k}{k^2}\int d^3x {\bf k}\times{\bf
 A(x)}e^{-i{\bf k}\cdot{\bf x}}\cdot\!\!\int d^3x^{\prime}{\bf
 k}\times{\bf B^{\ast}(x^{\prime})}e^{i{\bf k}\cdot{\bf
 x^{\prime}}}\\
   & = & \frac{1}{(2\pi)^6}\int\frac{d^3k}{k^2}\int
   d^3x{\bf A(x)}\times{\nabla}e^{-i{\bf k}\cdot{\bf
   x}}\!\!\cdot\!\!\int d^3x^{\prime}{\bf B^{\ast}(x^{\prime})}\!\times\!{\bf\nabla^{\prime}}e^{i{\bf k}\cdot{\bf
 x^{\prime}}}\\
   & = & \frac{1}{(2\pi)^6}\int\frac{d^3k}{k^2}\int
   d^3x{\bf\nabla}\!\times\!{\bf A(x)}e^{-i{\bf k}\cdot{\bf x}}\cdot\!\!\int d^3x^{\prime}
   {\bf\nabla^{\prime}}\!\times\!{\bf B^{\ast}(x^{\prime})}e^{i{\bf k}\cdot{\bf
 x^{\prime}}}\\
   & = & \int d^3x\int d^3x^{\prime}\frac{{\bf\nabla}\times{\bf
   A(x)}\cdot{\bf\nabla^{\prime}}\times{\bf
   B^{\ast}(x^{\prime})}}{4\pi\mid{\bf x}-{\bf x^{\prime}}\mid}.
   \label{sixteen}
 \end{eqnarray}
  Combining Eqs.\ (\ref{eight}),\ (\ref{twelve}) and \
  (\ref{sixteen}), we obtain our final result
  \begin{equation}
 \int d^3x\,{\bf A(x)}\cdot{\bf B^{\ast}(x)}=\int d^3x\int
 d^3x^{\prime}\frac{\left[{\bf\nabla}\cdot{\bf A(x)}{\bf\nabla^{\prime}}\cdot{\bf B^{\ast}(x^{\prime})}+
 {\bf\nabla}\times{\bf A(x)}\cdot{\bf\nabla^{\prime}}\times{\bf B^{\ast}(x^{\prime}})\right]}{4\pi\mid{\bf x}-{\bf
 x^{\prime}}\mid},
  \end{equation}
  recovering the Gubarev, Stodolsky and Zakharov result for the
  special case where the two fields are identical (and real).

  The same Fourier transform technique employed in this paper can
  be used to prove Helmholtz's theorem by which we mean the
  identity,
 \begin{equation}
{\bf V(x)}=\nabla\left(-\frac{1}{4\pi}\int
d^3x^{\prime}\frac{\nabla^{\prime}\cdot{\bf V(x^{\prime})}} {\mid
{\bf x}-{\bf x}^{\prime}\mid}\right)+\nabla\times\left(
\frac{1}{4\pi}\int d^3x^{\prime}\frac{\nabla^{\prime}\times {\bf
V(x^{\prime})}}{\mid {\bf x}- {\bf x}^{\prime}\mid}\right),
\end{equation}
demonstrating that a vector field ${\bf V(x)}$ is completely
determined, apart from an additive constant, once its divergence
and curl are specified. Indeed, the proof of either identity by
the above method provides an excellent illustration or student
exercise in the context of an undergraduate course in mathematical
physics, combining as it does the methods of Fourier analysis and
vector analysis.
  

\begin{thebibliography}{5}

\bibitem{gubarev}F. V. Gubarev, L. Stodolsky, and V. I. Zakarov,
``On the significance of the vector potential squared," Phys. Rev.
Lett. {\bf 86}(11), 2220-2222 (2001).

\bibitem{stewart}A. M. Stewart, ``On an identity for the volume
integral of the square of a vector field," Am. J. Phys. {\bf
75}(6), 561-564 (2007).

\bibitem{durand}L. Durand, ``Comment on `On an identity for the volume
integral of the square of a vector field,' by A. M. Stewart [Am.
J. Phys. 75 (6), 561-564 (2007)]," Am. J. Phys. {\bf 75}(6), 570
(2007).

\bibitem{generalization}After this note was submitted, it was
brought to the author's attention that in version (v3) of the
arXiv version of Ref. \onlinecite{stewart}
(http://au.arxiv.org/abs/0705.2081) Stewart has extended his
coordinate-space derivation of Eq.\ (\ref{one}) to the scalar
product of two different real vector fields.

\bibitem{complex}Since the two vector fields are quite arbitrary,
the use of the complex conjugate in Eq.\ (\ref{four}) and
thereafter is simply a matter of formal convenience.  But the
notation does serve as a reminder that the identity holds for
complex fields.

\end{thebibliography}
\end{document}